\documentstyle[12pt]{article}
\begin{document}
\renewcommand{\thefootnote}{\fnsymbol{footnote}}
\setcounter{page}{1}
\begin{titlepage}
\begin{flushright}
\large
INFN-FE 07-94 \\
July 1994
\end{flushright}
\vspace{3.cm}
\begin{center}
{\Large Determination of QCD condensates\\
 without hadronic spectra}\\
\vspace{0.5cm}
{\large Murman Margvelashvili${}^{(a,b)}$\footnote{E-mail:
 mm@infn.ferrara.it, mm@hepitu.kheta.georgia.su } and
Ketino Aladashvili${}^{(b)}$\footnote
{E-mail: keta@hepitu.kheta.georgia.su }}

\vspace{0.4cm}

{\em ${}^{(a)}$  INFN sezione di Ferrara, I-44100 Ferrara, Italy

\vspace{0.2cm}

${}^{(b)}$High Energy Physics Institute, Tbilisi State University \\
380086, Tbilisi, Georgia

 }

\vspace{0.3cm}

\end{center}
\vspace{1.0cm}
\begin{abstract}
The bounds on the values of gluon, four-quark and quark-gluon condensates
are derived from the requirement of consistency of the sum rules for various
correlators of the hybrid current $a_{\mu}=g\bar d\gamma_{\rho}\tilde
G_{\rho\mu}u$. The upper bound for the gluon condensate is found to be less
then twice its standard value  while for the four-quark condensates the
violation of factorization by a factor 3-4 is allowed.
The value of the quark-gluon condensate is given as a function of
$\left<\frac{\alpha_s}{\pi} G^2\right>$ and $\left<(\bar q\Gamma q)^2\right >$
which for the standard values yields $m_0^2=0.63~\mbox{GeV}^2$ .
Our procedure of simultaneous solution of a number of sum rules
allows the quantitative error analysis and is appropriate
for condensate determination also in other cases.
\end{abstract}
\end{titlepage}
\renewcommand{\thefootnote}{\arabic{footnote}}
\setcounter{footnote}{0}
\newpage
\begin{section}*{Introduction}
QCD vacuum condensates, the vacuum expectation values of local operators,
introduced in ref.\cite{svz} as essential ingredients of the QCD sum
rules (SR) method, are the fundamental parameters of theory. They
parametrize the properties of QCD vacuum and therefore have relation
to almost all aspects of the hadron physics.

Since the introduction of QCD condensates, considerable effort has been
devoted to determination of their values, which however up to now
has given a little effect. The existing
theoretical schemes provide only qualitative estimates,
while semiphenomenological analyses on the basis of QCD SR approach
have resulted in unacceptably wide ranges of possible values, even
for the basic condensates of lowest dimensions.

The most commonly used values of QCD condensates are the so called
"standard values" (SV) of ref.\cite{svz}. For the gluon condensate this is

\begin{equation}
\left<\frac{\alpha_s}{\pi}
G^2_{\mu\nu}\right>_{st}=(330~\mbox{MeV})^4
\end{equation}
as derived from charmonium phenomenology, while for the four-quark condensate -
the "factorized" value
\begin{equation}
\left<(\bar q\Gamma q)^2\right >_{fact}=
\frac{1}{N(\Gamma)}<\bar qq>^2
\end{equation}
whith the quark condensate given by
$<\sqrt{\alpha}\bar qq>=(-240~\mbox{MeV})^3$, and $N(\Gamma)$ -
a normalization factor depending on the matrix $\Gamma$.

The mixed quark gluon condensate has been
estimated from the baryon sum rules \cite{io} to be
\begin{equation}
\left<ig\bar q\sigma_{\mu\nu}G^a_{\mu\nu}\frac{t^a}{2}q\right>=
m_0^2<\bar qq>
\end{equation}
with $m_0^2=0.8~\mbox{GeV}^2.$

In a great number of articles the different types of  sum rules (Borel SR
\cite{svz}, FESR\cite {fesr}, Gauss SR \cite{gaus} etc.)
have been applied for extraction of the actual condensate values
from experimental data \cite{ei,we} (here we quote only a few).
Most of them use
the measured hadronic spectra from $e^+e^-$ annihilation or $\tau$
decay processes. These studies however have not led to a consistent
picture. The resulting values of condensates considerably differ
from each other, most of them  exceed the SV several
times. For the gluon condensate the corresponding factor varies in the
range (1-6) while for the four-quark condensate it is up to an order of
magnitude. As for the parameter $m_0^2$, the estimates
ranging from $0.2$ to $1.1~\mbox{GeV}^2$ can be found in the literature.

On the other hand there are still arguments against such strong
deviations of the vacuum condensates from their SVs.
So in ref.\cite{nsvz} the restrictions on the value of gluon condensate
have been obtained from the requirement of consistency of different SR
for  the correlators of a hybrid current
$a_{\mu}=g\bar d\gamma_{\rho}\tilde G_{\rho\mu}u$.
The two point functions
$<0|T(a_{\mu},a_{\nu})|0>$ and $<0|T(\bar u\gamma_{\mu}\gamma_5 d,a_{\nu}|0>$
have been considered and the gluon condensate has been bounded by the
requirement that these different sum rules yield the same result for the
pion matrix element $<0|a_{\mu}|\pi(p)>=-if_{\pi}\delta^2p_{\mu}$.
Using the factorization assumption for the four-quark condensate, the
authors have concluded that $<\frac{\alpha}{\pi}G^2_{\mu\nu}>$ is forbidden to
exceed its standard  value by more then some 40\%.
The value of  the quark-gluon condensate has been
calculated in \cite{ovch} by means the sum rules for the correlator of
$a_{\mu}$ with  the pseudoscalar current.
The authors, using the SV for the four-quark
and the gluon condensates together with $\delta^2=0.2$ from
ref.\cite{nsvz}, have arrived at the value of eq.(3).

The mentioned uncertainties of the condensate values lead to
considerable ambiguities in the predictions of QCD SR as well as
in various other applications. They also arise doubts in the consistency of
description of the vacuum properties by means of local condensates.
However considering the situation with condensate
determination
one has to remember that QCD SR is a semiquantitative method and
without well defined error analysis the results have
quite a limited
value. Unfotunately not all of the attempts of condensate determination
meet this requirments. The intrinsic uncertainties of the method
related to the truncation of power series and to the imperfection of the
model spectra are not even parametrized in a fully satisfactory
way. Hence different types of SR, influenced by these uncertainties
in different ways, produce the results which seem to be
hardly compatible with each other. Presently the systematic study
of different sum rules with clear statements about related errors
is highly desirable; This will allow either to constrain the
values of the basic condensates to some reasonable ranges or
to show the limits of consistency of the whole approach.

In the present paper we perform the simultaneous analysis of the
sum rules for different correlators of the current $a_{\mu}$
in order to impose bounds on the values of the gluon
quark-gluon and the four-quark condensates.
We are motivated by several reasons. First of all let us note
that in ref.\cite{nsvz} the gluon condensate has been  fixed by
assuming the SV for the four-quark condensates.
On the other hand we have quite a reliable estimate based on the
$\tau$ decay data and low energy theorems of current algebra,
showing  that for the definite four-quark condensate the
factorization approximation is violated by a factor of (2.5-3)
\cite{we}. So it is desirable to find bounds on the above
parameters without any ad hoc. assumptiontions. It is also
interesting to find the allowed region for $\delta^2$ which is
an important parameter determining many dynamical characteristics
of pion. Our motivation is also to present the framework which
allows the quantitative error analysis and is more adequate for a
systematic study of a number of different  sum rules.

Our suggestion briefly is the following.
Instead of the standard procedure we propose to consider the sum
rules at different values of the Borel parameter $M^2$ as
independent equations with estimated errors. Then we find the
minimal $\chi^2$ solution of the system of such equations and
test its stability against variations of different parameters
involved.
In section 2 we formulate the problem and present the equations
used for condensate determination. In section 3 we describe the
procedure for solution of a system of sum rules. Sect.4 contains
the presentation of results,  discussion and comments.

\end{section}

\begin{section}*{Equations}
In order to impose the bounds on the possible values of different
condensates and $\delta^2$
we consider the following set of two point functions of the hybrid current
$a_{\mu}$:

\begin{eqnarray}
i\int dx e^{iqx}<0|T\left(\bar u\gamma_{\alpha}\tilde G_{\alpha\mu}d(x),
\bar d\gamma_{\beta}\tilde G_{\beta\nu}u(0) \right)|0>  =
g_{\mu\nu}\Pi^{d}_1(q^2)+q_\mu q_\nu \Pi^{d}_2(q^2)  \nonumber \\
i\int dx e^{iqx}<0|T\left(\bar u\gamma_{\mu}\gamma_5 d(x),\bar d\gamma_{\rho}
\tilde G_{\rho\nu}u(0) \right)|0> = g_{\mu\nu}\Pi^{ax}_1(q^2)+q_\mu q_\nu
\Pi^{ax}_2(q^2)   \\
i\int dx e^{iqx}<0|T\left(\bar u \gamma_5 d(x),\bar d\gamma_{\rho}
\tilde G_{\rho\mu}u(0) \right)|0> =  q_\mu  \Pi^{ps}(q^2)  \nonumber
\end{eqnarray}

There are several reasons which make this correlators appropriate for
condensate determination. First of all, their perturbative expansions
start at the two loop level, so the nonperturbative terms become dominant
 and hense easier to determine. In contrast to the vector and axial-vector
current correlators, in the operator expansion (OPE) of the two point
functions (4) the correlation of gluon and four-quark
condensates values is negative, which allows to bound their absolute
values from above.
The number of unknown parameters in the sum
rules is reduced by the fact that imaginary parts of
this correlators are contributed by the same physical states.
These are the pion, $a_1$ and higher mass states which in the following
will be attributed to the continuum.
In the sum rules we use the usual model
spectrum   with $\pi$ and $a_1$ as narrow resonances and the
continuum equal to
the perturbative one, starting at some point $s_0$. After standard
manipulations one obtains the following
SRs for $\Pi_2^d, \Pi_2^{ax}  \Pi_1^{ax}$ and $\Pi^{ps}$ correspondingly:

\begin{eqnarray}
\lefteqn{ \frac{1}{72} \left<\frac{\alpha_s}{\pi} G^2\right>_{st}G+\frac{8}{9}
\frac{\left<\sqrt{\alpha_s}\bar qq \right>^2_{fac}Q}{M^2}  =}  \nonumber \\
 & & =  \frac{f_{\pi}^2 \delta^4}{M^2}+
\frac{f_{a}^2\delta^4_a}{M^2}\exp(-m^2_a/M^2)
-\frac{\alpha_s}{80\pi^3}M^4\left( 1-(1+x_0+\frac{x_0^2}{2})e^{-x_0}\right)
\end{eqnarray}
\begin{eqnarray}
\lefteqn{\frac{1}{6M^2} \left<\frac{\alpha_s}{\pi} G^2\right>_{st}G+
\frac{64\pi}{27}
\frac{\left<\sqrt{\alpha_s}\bar qq \right>^2_{fac}Q}{M^4} =}  \nonumber \\
& &  = \frac{f_{\pi}^2\delta^2}{M^2}-
\frac{f_{a}^2\delta^2_a}{M^2}\exp(-m^2_a/M^2)
-\frac{\alpha_s}{18\pi^3}M^2\left( 1-(1+x_0)e^{-x_0}\right)
\end{eqnarray}
\begin{eqnarray}
\lefteqn{\frac{64\pi}{27} \frac{\left<\sqrt{\alpha_s}\bar qq \right>^2_{fac}
Q}{M^2}}  \nonumber \\
& &   =\frac{f_a^2 m_a^2\delta^2_a}{M^2}\exp(-m^2_a/M^2)
-\frac{\alpha_s}{18\pi^3}M^4\left( 1-(1+x_0+\frac{x_0^2}{2})e^{-x_0}\right)
\end{eqnarray}
\begin{eqnarray}
\frac{m_0^2 \left<\bar qq \right>}{M^2}+\frac{\pi^2}{9}
\frac{\left<\frac{\alpha_s}{\pi} G^2\right>\left<\bar qq\right >}{M^4}
=\frac{\delta^2 \left<\bar qq \right >}{M^2}-\frac{\alpha_s \left< \bar qq
\right >}{3\pi}(1-e^{-x_0'})
\end{eqnarray}
where $f_{\pi}=133~\mbox{MeV}$ and $f_a=170~\mbox{MeV}$\cite{svz} are the
leptonic decay constants  of $\pi$ and $a_1$  mesons correspondingly.
$m_a=1260~\mbox{MeV}$ is the  $a_1$ mass and in an analogy with the pion
matrix element we have
$\epsilon_{\mu}f_a\delta^2_a=<0|a_{\mu}|a_1>$. We have introduced the
coefficients
\[G \equiv \left<\frac{\alpha_s}{\pi} G^2 \right> /\left<\frac{\alpha_s}{\pi}
G^2 \right>_{st}
{}~~~~~~~~Q\equiv \left< (\bar q\Gamma q)^2 \right> /
\left<(\bar q\Gamma q )^2\right>_{fac}\]
measuring the deviations of corresponding condensates from their standard
values. Note that Q is some average coefficient for different four-quark
condensates which appear in the operator product expansion (OPE) of the
considered correlators. We have
$x_0=\frac{s_0}{M^2}$ where $s_0$ is the continuum onset in eqs.(5-7).
On the orther hand in eq.(8) where there is no $a_1$ contribution the
corresponding continuumthreshold is taken to be lower
$s_0'<s_0$ . We disregard the anomalous
dimensions of the operators and consider the corresponding matrix elements
at $\mu^2\approx M^2$. All the terms except the unknown condensate
contributions are moved to the right hand sides of the corresponding
equations.

A few comments are in order before we proceed further.
Equations (5) and (6) are taken from \cite{nsvz}. In eq.(7) we have
corrected the sign of the second term in the rhs. which was incorrect
in ref.\cite{ovch}. This change reduces the resulting value of
quark-gluon condensate considerably.
The SR (8) is not very reliable since its different sides show different
$M^2$ behavior, however we have added it to the system just to illustrate
some points of our approach.
The second term of the lhs. of (7) is obtained by the factorization of
dimension 7 quark-gluon condensate, however due to numerical
smallness of this term the approximation has practically no effect
on the results.

Now let us recall the standard sum rules
procedure in order to understand better our proposal.
Note that QCD sum rules are approximate equations
which depend on the borel parameter $M^2$. The errors in this equations
are mainly caused by two independent reasons. At low $M^2$ the
inaccuracy of equations is mostly caused by the errors resulting from
the truncation of OPE series and from the uncertainties of theoretical
parameters. Whereas at high values of borel parameter the errors caused
by inaccuracies of experimental (or model) spectra become more essential.
At the intermeidate values of the Borel parameter there
is some interplay of both of these errors.
If there is a region of $M^2$ where both of them are
moderate ("fiducial region"), then
the physical quantity in question is extracted  by equating
theoretical and experimental parts of SR, as well as their derivatives,
at some particular point within this region.  The reliability
of the result is then tested by its stability against variation of $M^2$.

The above procedure is sufficient for deriving qualitative results
and estimates, however it contains considerable amount of arbitrariness
related to the choice of fiducial region, determination of the
central  value and erros of the final result. Besides it is difficult to
compare the results obtained in different sum rules by means of such a
procedure. The well established quantitative results can be obtained
by QCD sum rules method
only in case if these uncertainties will be  somehow eliminated.

Instead of the traditional treatment we suggest to consider the Borel
transformed SR (5-8) at different $M^2$ as independent approximate
equations in which the estimates of the expected errors and their
correlations should be done. Then the system of such equations can
be solved in the sense of minimal $\chi^2$, with account of this errors.
This provides a more natural concept of similarity of different
parts of the sum rules than the standard procedure.
Calculating $\chi^2$ per degree of freeedom we have the measure of
consistency of different sum rules entering the system.
 Then
we scan all the possible values of the unknown parameters $\delta^2$
and $\delta^2_a$ and find the region in the three dimensional space
of parameters $G$, $Q$ and $m_0$ for which the system is consistent.
This gives us the bounds on the possible values of condensates
as well as correlations among them.

Of course such a way is related to many subtleties
of the error analysis, however this is unavoidable if one wants
to controll the precision of the results obtaineds by the sum rules method.
At least moving in this direction we will investigate the capability of
the method to provide accurate numerical results.

\end{section}

\begin{section}*{Condensate determination}

In order to determine the allowed ranges for the gluon, four-quark
and quark-gluon condensates we perform the following steps:

{\bf 1. Construction of the system of equations}.
Taking some values of the parameters $\delta^2$ and $\delta^2_a$ we
evaluate the equations  (5-8) at four different values of the borel paremeter
$M^2_{\alpha}$
$\alpha=1,.,4$ and build up a system of 16 linear equations for determination
of the parameters Q, G and  $m_0^2$. We take the $M^2_{\alpha}$ points in
the  range $0.6\div 2~\mbox{GeV}^2$, however they should not be too close
to each other  in order to avoid strong error correlations among the
resulting equations (see below).

{\bf 2. The estimate of expected errors.}
To set the scale of expected errors in the resulting equations we first
check separately the SRs (5-8) and adopt some relative errors $w_i$ for
each of them at $M^2=1~\mbox{GeV}^2$. The eqs. (5-7) are easily stabilized
by appropriate choice of $s_0$ for reasonable values of
condensates and $\delta^2$. Hence, we have no reason to expect that some
important contributions are missing. On the other hand eq.(8) exhibits
different $M^2$ behavior of its right and left hand sides and is
clearely less reliable. So, we should take $w_8>w_{5,6,7}.$
As initial guess we assume $w_{5,6,7}=0.1$  and $w_8=0.3 $ i.e.
10\% relative error at $M^2=1~\mbox{GeV}^2$ for eqs.(5-7) and 30\% of that
for (8). We require that the results shold be stable against reasonable
variations of these numbers.

The $M^2$ dependence of errors is taken to be common for all the considered
SRs and is discribed by the error distribution function

\begin{eqnarray} \label{D}
      D(M^2)=\left\{ \begin{array}{ll}
	\frac{1-M_0^2}{M^2-M_0^2} & \mbox{if $M^2 < 1~\mbox{GeV}^2$}\\
	1 & \mbox{if $M^2\geq1~\mbox{GeV}^2$}
	\end{array}
	\right.
\end{eqnarray}

Such a choice of $D(M^2)$ can be justified in the following way;
It is known that at low $M^2$ the higher order power corrections
in the OPE series become essential and at some $M^2$ the expansion
blows up. So, we assume the pole like behaviour at $M^2<1~\mbox{GeV}^2$
with $M^2_0$ being some effective convergence radius of the series.
On the other hand for the higher $M^2$ values the contributions of $a_1$
and higher states, which become the main source of the errors, remain
practically constant, so we assume that the error distribution to is
uniform for $M^2>1~\mbox{GeV}^2$. Note that the absolute values of SRs
decrease with $M^2$ so the above of $D(M^2)$ corresponds to increasing
{\em relative} error both for high and low $M^2$.

Finally, the expected {\em absolute} errors for the equations of
our system are calculated as

\begin{eqnarray}
\Delta_i(M^2_{\alpha})=\omega_i D(M^2_{\alpha})A_i(M^2_{\alpha})
\end{eqnarray}
where $A_i(M^2)$ denotes the rhs.s of (5-8).

{\bf 3.Solution of the system of equations}.
To find the allowed regions of the condensate values
we take different values of the parameters $\delta^2$ and $\delta_a^2$
and for every combination we find the solution for Q,G and $m_0$ at
which the system is maximally consistent. Such a way is dictated by
simplicity reasons, since the system is linear in condensates.
As a measure of consistency of the system we use the minimal $\chi^2$
criterium the use of which we will now try to advocate.

It is clear, that the errors of our equations are not statistical and
should be somehow correlated with each other. However these correlations
depend on many different factors, like unknown higher order terms in
OPE, perturbative corrections, discrepancies between model and real spectra,
possible non OPE
contributions etc. The knowledge of all these factors would be almost
equivalent to the solution of QCD for our case, and we are clearly far
from this.
The best that we can do at present, is to consider the errors of different
SRs as uncorrelated; As for errors within each of the sum rules (5-8),
we have to keep the $M^2$ points sufficiently separated to ensure that
this correlations are small.
The effect of our assumption will be less important for a greater number
of equations,
and can be partly compensated by increasing the error bars.
In other words we suggest to treat the uncertainties caused by a great
number of unknown parameters as some statistical errors.
So, assuming that in each of 16 equations the errors are independent
we sum them quadratically to calculate $\chi^2$  per degree of freedom.

We scan all the possible values of the parameters $\delta^2$ and
$\delta^2_a$ and for each pair
find the corresponding values of $G$, $Q$ and $m_0^2$ with minimal
$\chi^2_{d.f.}$. If the resulting  $\chi^2_{min}<1$ then the system
is considered to be consistent and the values of the above parameters -
allowed. Thus we obtain the allowed region in the three dimensional space
of parameters
$G$, $Q$ and $m_0^2$. The corresponding region in the plain of $(G,Q)$ is
shown in fig.1. The inner curve - $a)$ corresponds to $\chi^2_{df.min}= 1$
However to be less stringent we allow also solutions with
$\chi^2_{df.min} \leq 2$ (bounded by the curve $b)$)which corresponds to
$\approx 15\%$ relative error at $M^2=1~\mbox{GeV}^2$ in eqs.(5-7) .

One should keep in mind that for each  $\delta^2$ and $\delta_a^2$ the
parameters $G$, $Q$ and $m_0^2$, giving minimal $\chi^2_{df}$ solution, have
correlated errors which have also to be taken into account. For a few
solutions this is illustrated by ellipses in fig.2 . These ellipses, for
every solution with $\chi^2_{min}$, include
the points for which  $\chi^2 \leq \chi^2_{min} +1$.
The area covered by all these ellipses is similar to
the area with $\chi^2_{min} \leq 2$ bounded by the curve b) of fig.1.

{\bf 4.Stability check and parameter adjustment.}
The results should be
stable against the variations
of the parameters involved in the derivation.
So we check the stability of the boundaries and individual solutions,
and adjust the parametersl in order to obtain the maximal
stability of the allowed region.

If the error distribution function is realistic, then the results
should not depend considerably on the variation of the endpoints
$M^2_1$ and $M^2_4$.
The error distribution function (9) with $M_0^2=0.45~\mbox{GeV}^2$
just satisfies this requirement. The
boundary of the allowed region remains practically unchanged
when $M^2_1$ and $M^2_4$ vary in the intervals  $(0.6-0.9)~\mbox{GeV}^2$
and $(1.6-2)~\mbox{GeV}^2$ respectively. The region shown in fig.1 is
obtained with $M^2_{\alpha}$ values taken at 0.7,1.1, 1.4,1.9 $~\mbox{GeV}^2$.
For comparison in fig.3 we have also plotted dependence of the allowed
$(G,Q)$ region on the lower endpoint $M^2_1$ for $M_0^2=0.3~\mbox{GeV}^2$.

The final result is also insensitive to individual variations of
$w_i$ within about a factor of 2. However, this is not so for $w_8$
and the allowed region feels the erroneous eq.(8) unless we take
$w_8\geq 0.5$.
Our final result is not strongly affected by the considerable variations
of $s_0$ and $s_0' $ as can be seen from fig.4. This however is hardly
surprising in view of the two loop suppression of perturbative
contributions.
\end{section}

\begin{section}*{Results and discussion}

Thus, we have found the allowed regions for the gluon, four-quark and
quark-gluon condensates as well as the parameters $\delta^2$ and
$\delta_A^2$. We conclude that the system of sum rules (5-8) is
consistent with the standard values of condensates.
{}From fig.s 1,2 one can see that the gluon condensate
can differ from its SV at most by a factor of 2 and this
happens if the four-quark condensates are close to their factorized
values. On the other hand the latter are less tied to their SV and
can exceed them 3-4 times.
The upper bounds on the values of condensates are more strict
than that derived from vector and axial channels. This is due to the
fact that in  the considered equations both the gluon and the four-quark
condensates have the
same sign and thus cannot compensate the growth of each other.
Note however that we did not
distinguish different types of the four-quark condensates.

Since for every $\delta$ and $\delta^2_a$ we obtain a triplet
of parameters $G$, $Q$ and $m_0^2$, we thus have some functional
dependence among their possible values, described by two dimensional
surface in the three-dimentional space of this parameters.
This dependence is almost linear and can be
expressed by the formula: \[m_0^2=(0.24G+0.3Q+0.09) ~\mbox{GeV}^2\]
which for the SV of condensates (i.e. G=Q=1) yields:
\[ m_0=0.63~\mbox{GeV}^2 \] in difference with eq.(3)

The allowed regions of condensates correspond to the following ranges
of the pion
and $a_1$ matrix elements: $0.1~\mbox{MeV}^2<\delta^2<0.3~\mbox{MeV}^2$ and
$0<\delta^2_a<0.2~\mbox{MeV}^2$ which are thus also bounded by the
requirement of consistency of the system of SR.

The proposed procedure of analysis can be easily extended to other systems
of two point functions
and to a greater number of equations, however in our opinion it is already
useful for analysis of a single SR, since it allows clear and testable
assumptions about errors and their distribution. We hope, that this will
be helpful for further restriction of condensate values or establishing
the limits of applicability of the QCD SR method.
\end{section}

\begin{section}*{Acnowledgements}
We thank I.Avaliani, Z.Berezhiani, J.Gegelia, A.A.Ovchinnikov,
M.Shaposhnikov for helpful discussions.
We are especially thankful to V.Kartvelishvili for his kind assistance
in our work.

One of the authors (M.M.) wants to express his gratitude to
Z.Berezhiani and G.Fiorentini for their warm hospitality at Ferrara section of
INFN where the part of this work was done.

The work of M.M. was in part supported by the grant of International Science
Foundation.
\end{section}


\newpage
\begin{center}
{\Large \bf Figure captions}
\end{center}

\vspace {3.cm}

{\bf Fig.1} The allowed region in the plane $(G, Q)$ calculated
for $M^2_i=0.7, 1.1, 1.4, 1.9~\mbox{GeV}^2$. The curve {\bf a)} bounds the
region
with $\chi^2_{min} \leq 1$ assuming 10\% rel.error in eqs.(5-7) at
$M^2=1~\mbox{GeV}^2$.
{\bf b)} bounds the region with $\chi^2_{min} \leq 2$. \\

\vspace {2.cm}

{\bf Fig.2} The allowed region of $(G,Q)$ plane taking into account the
errors of the solutions with $\chi^2_{min}\leq 1$ (inside curve a) of fig.1).
The ellipses indicate the areas with
$\chi^2\leq \chi^2_{min}+ 1$ for three different solutions, one of them
corresponding to the SVs. \\

\vspace {2.cm}

{\bf Fig.3} Allowed $(G,Q)$ regions with $\chi^2_{min}\leq 1$ calculated using
the error distribution function (9) with $M^2_0=0.3~\mbox{GeV}^2$;
a)for $M^2_1=0.6~\mbox{GeV}^2$
b)for $M^2_1=0.8~\mbox{GeV}^2$. \\

\vspace {2.cm}

{\bf Fig.4} Allowed regions with $\chi^2_{min}\leq 1$ for different choice of
$s_0$ and $s_0'$. \\ $a)$ - $s_0=2~\mbox{GeV}^2$ $s_0'=1.5~\mbox{GeV}^2$
$b)$ - $s_0=3~\mbox{GeV}^2$ $s_0'=2.5~\mbox{GeV}^2$

\end{document}